\documentclass[12pt]{iopart}
\usepackage{comment}
\usepackage{xcolor}
\usepackage{graphicx}
\usepackage{dcolumn}
\usepackage{bm}
\usepackage{caption}
\usepackage{hyperref}
\usepackage{amssymb}
\usepackage{float}
\usepackage{booktabs}

\expandafter\let\csname equation*\endcsname\relax
\expandafter\let\csname endequation*\endcsname\relax
\usepackage[version=4]{mhchem}
\usepackage{soul}
\usepackage{siunitx}
\DeclareSIUnit\barn{b}

\begin{document}

\title{Trapping $\mathbf{Ba}^+$ with Seven-fold Enhanced Efficiency Utilizing an Autoionizing Resonance}

\author{Noah Greenberg, Brendan M. White, Pei Jiang Low, and Crystal Senko}
\address{Institute for Quantum Computing and Department of Physics $\&$ Astronomy, University of Waterloo, Waterloo, Ontario, N2L 3G1, Canada}

\pagestyle{plain} 

\ead{n2greenberg@uwaterloo.ca}
\vspace{10pt}

\begin{abstract}


Trapped ions have emerged as a front runner in quantum information processing due to their identical nature, all-to-all connectivity, and high fidelity quantum operations.
As current trapped ion technologies are scaled, it will be important to improve the efficiency of loading ions, which is currently the slowest process in operating a trapped ion quantum computer.
Here, we compare two isotope-selective photoionization schemes for loading $^{138}\mathrm{Ba}^+$ ions.
We show that a two-step photoionization scheme ending in an autoionizing transition increases the ion loading rate nearly an order of magnitude compared to an established technique which does not excite an autoionizing state.
The only additional technology required to implement the autoionizing transition is a commercial diode laser.
Our technique can be extended to all isotopes of barium, and autoionizing resonances exist in every species currently used for trapped ion quantum processing, making this a promising technique to drastically increase the loading rates for all trapped ion computers.


\end{abstract}

%
%
%
%
%

\section{Introduction}
Barium is a versatile quantum information carrier with the capacity to be used as an optical, metastable, or ground-state qubit \cite{omg}. 
Additionally, barium has been demonstrated as a \textit{qudit}, controlling and reading out up to 13-levels \cite{low2023control} by taking advantage of its unusually long-lived metastable $\ce{5d\:{}^2D_{5/2}}$ state \cite{PhysRevA.41.2621}. 
The flexibility of the atomic structure in barium and the visible wavelengths used to drive the electric-dipole transitions are also attractive from a practical perspective \cite{binaimotlagh2023guided}.
All of these key features of barium, coupled with extraordinary experimental state-preparation and measurement results \cite{Hucul2017, Christensen2020_thesis, PhysRevLett.129.130501} has captured the attention of researchers and motivated the trapping of barium in the most sophisticated surface traps as a front-running platform for quantum information processing \cite{shi2023ablation, Graham2014}. 
Unfortunately, a caveat of working with metallic barium is that it can oxidize within seconds when exposed to atmosphere, and specifically, the $^{133}\mathrm{Ba}^+$ isotope can only be sourced as a salt and used in microgram quantities. 
For these reasons, the focus of this work is on maximizing the loading rate of barium to quickly generate long chains of ions for quantum information processing utilizing laser ablation. 


Typically, photoionization for trapping ions is done using resonance-enhanced multi-photon ionization \cite{ASHFOLD19991424}. 
This can involve a resonant ``first step'', using a laser that drives an electronic transition in the neutral atom, along with a ``second step'' using light that is sufficiently energetic to bridge the electron from the excited state to the unbound continuum. 
Isotope shifts in the first step transition are generally large enough ($> 10$ MHz) so that only the target isotope of neutral atoms are excited and then ionized. 
Researchers have employed a variety of different first steps in two-step photoionization schemes for the isotope-selective loading of barium.
Some relevant parameters for these first step transitions are outlined in Table \ref{table:FirstStep}.  


\begin{table}[H]
\begin{center}
    \begin{tabular}{c|ccc}
    \toprule\toprule
    $\lambda$ (nm) & Intermediate State & Linewidth (MHz) & \ce{6s^2\:{}^1S_0} Branching Ratio \\ \hline
    413& \ce{5d 6p\:{}^3D_1} & $9.15 \pm 0.26$ \cite{Niggli1987} & 0.026(13) \cite{Niggli1987} \\ 
    554& \ce{6s 6p\:{}^1P_1} & $19.02 \pm 0.19$ \cite{Niggli1987} & 0.9966(0.2) \cite{Niggli1987} \\ 
    791& \ce{6s 6p\:{}^3P_1} & $0.820 \pm 0.050$ \cite{AKULSHIN199254} & 0.38 \cite{PhysRevA.61.062509} \\ \bottomrule\bottomrule
    \end{tabular}
    \caption{
    Comparison of first steps used to drive a transition in neutral barium between the ground state, $\ce{6s^2\:{}^1S_0}$, and an excited intermediate state, as part of a two-step photoionization process for loading $\mathrm{Ba}^+$.
    The branching ratio represents the likelihood that an atom in the intermediate state will decay back to the ground state.
    In this work we use $\qty{554}{\nano\meter}$ as the first step to drive a strong dipole-allowed transition, with a branching ratio almost entirely back down to the ground state.
    }
    \label{table:FirstStep}
\end{center}
\end{table}


Utilizing an autoionizing resonance for the second step can significantly enhance the photoionization cross-section \cite{PhysRev.38.873, vant2006photoionization}.
By contrast, previous work loading ions with photoionization has typically used a non-resonant second step (\textit{i.e.} not ending in an autoionizing transition).
The non-resonant photoionization scheme ionizes an electron directly into the continuum, while the autoionizing scheme uses a resonant transition to a quasi-bound state embedded in the continuum, providing roughly an order of magnitude higher experimentally measured photoionization cross-section.
The probability of trapping is proportional to the photoionization cross-section $P_T \propto \sigma_P$ \cite{PhysRevA.74.063421}, so by using a more efficient photoionization scheme, the probability of trapping can be increased substantially. 

Autoionizing resonances can be found in the ionization spectrum of all elements with two valence electrons, including barium.
These resonances, which are found past the first ionization threshold, describe states where the wavefunctions of the electrons have become correlated \cite{UFano_1983}. 
They are the result of interference between a discrete state and the ionization continuum \cite{PhysRev.124.1866}, and they are found as distinct peaks in the photoionization cross-section. 
Typically, the resonance is characterized as a Fano-profile \cite{BWillke_1993}, allowing useful information to be extracted from the measured cross-section. 
The energies and cross sections of these resonances can be calculated with multi-channel quantum defect theory analysis \cite{M_Aymar_1990, MAymar_1979, RevModPhys.68.1015} or an eigenchannel R-matrix method \cite{PhysRevA.42.5773, PhysRevA.44.1773}.
The cross-sections have also been empirically measured \cite{BWillke_1993, PhysRevLett.67.2131, Kalyar}.


If the photon energy $\hbar\omega$ of the second step is tuned to a known autoionizing resonance, then the two valence electrons will become doubly excited and enter a quasi-bound state $\mathrm{Ba}^{**}$ embedded in the ionization continuum with an incredibly short lifetime, $\tau \ll \qty{1}{\nano\second}$:
\begin{equation}
    \hbar\omega + \mathrm{Ba}(\ce{6s 6p\:{}^1P_1}) \rightarrow \mathrm{Ba}^{**} \rightarrow \mathrm{Ba}^+ + \mathrm{e}^-.
\end{equation}

In this manuscript, two different photoionization schemes are compared for isotope-selective loading of $^{138}\mathrm{Ba}^+$ ions. 
Each scheme is a two-step process with the first step using $\qty{554}{\nano\meter}$ light to drive a strong dipole transition between the $\ce{6s^2\:{}^1S_0} \leftrightarrow \ce{6s 6p\:{}^1P_1}$ states. 
The second step for ionization uses either a non-resonant photoionization scheme with $\qty{405}{\nano\meter}$ light, or an autoionizing scheme near $\qty{390}{\nano\meter}$. 
The two schemes are presented in Fig. \ref{fig:Fig1}a.
Nearly an order of magnitude increase in loading rate for $^{138}\mathrm{Ba}^+$ ions is observed when using the autoionizing scheme compared with the non-resonant scheme.
Furthermore, the autoionizing scheme requires significantly lower amounts of optical power in the second step to trap long chains of $^{138}\mathrm{Ba}^+$, which reduces the likelihood of trap charging. 
The observed increase in loading rates is consistent with the increase in the previous experimentally measured photoionization cross-sections of the two compared photoionization schemes \cite{BWillke_1993, PhysRevLett.67.2131, Kalyar} . 

\section{Experimental Overview}

\begin{figure*}
    \centering
    \includegraphics[width=\linewidth]{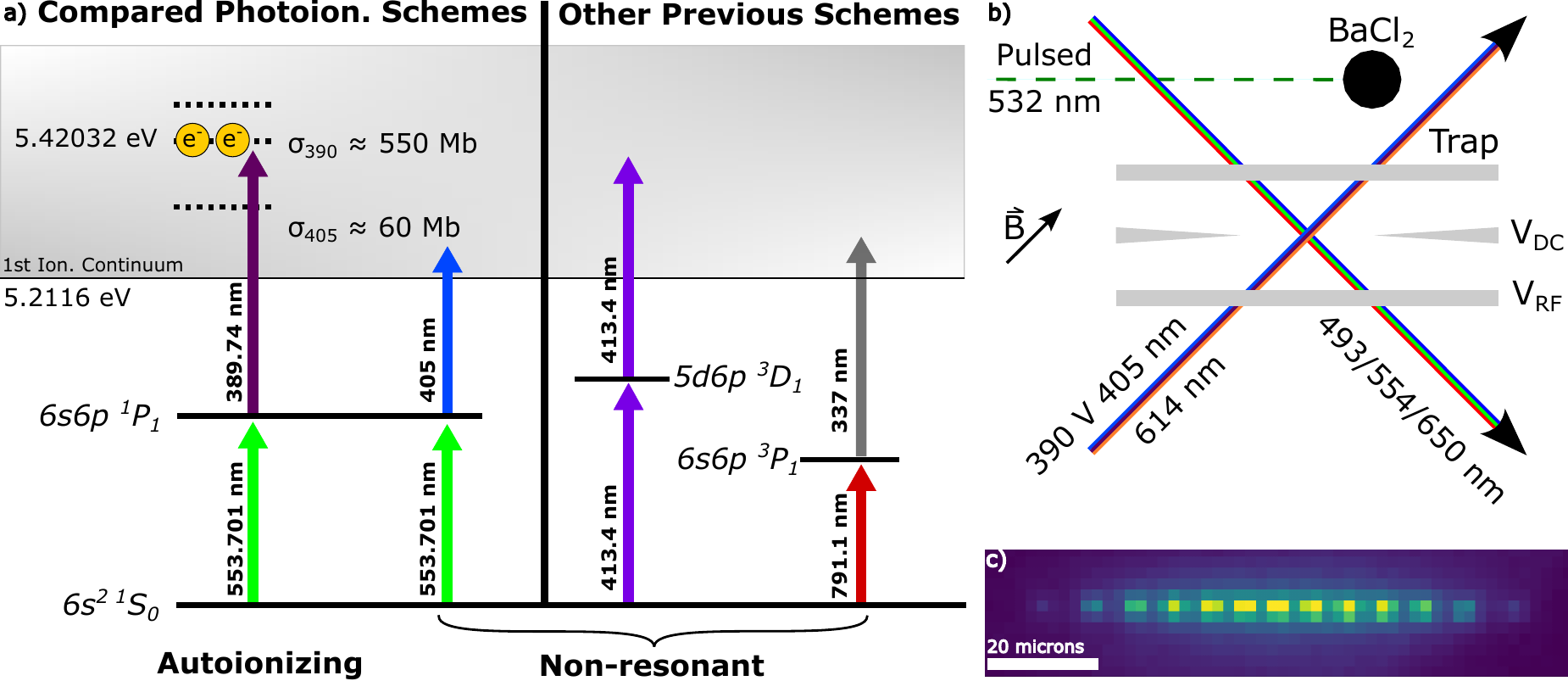}
    \caption{
    a) Comparison of the two different photoionization schemes in the text, with the previously measured cross-sections $\sigma_{390}$ and $\sigma_{405}$ \cite{BWillke_1993}. 
    Each scheme uses the exact same first step, exciting neutral barium to the $\ce{6s 6p\:{}^1P_1}$ state. 
    The second step varies depending on the scheme, either ending in an autoionizing resonance or somewhere in the continuum not resonant with a specific transition. Autoionizing states are represented by dashed lines in the continuum.
    b) Experimental overview of the beam geometry.
    The $\qty{554}{\nano\meter}$ beam is orthogonal to the atomic plume to reduce Doppler broadening and also orthogonal to the second ionization beam (either 390 or $\qty{405}{\nano\meter}$) to reduce the ionization volume. 
    c) A chain of thirteen $^{138}\mathrm{Ba}^+$ ions trapped with a single ablation pulse using the autoionizing scheme. 
    The outer ions sit far from the center of the beam, causing them to be dim compared to the center ions. 
    Each pixel is $\approx \qty{2}{\micro\meter}$.
    }
    \label{fig:Fig1}
\end{figure*}

The comparison of the two different photoionization schemes was conducted on a four-rod trap using IonControl software \cite{osti_1326630}. 
These experiments are performed using laser ablation on a natural abundance, stable, salt ($\mathrm{BaCl}_2$) target in an isotope-selective manner with a two-step photoionization process to trap $^{138}\mathrm{Ba}^+$ \cite{PhysRevA.105.033102}.
As depicted in Fig. \ref{fig:Fig1}b, the trap is comprised of four electropolished tungsten rods with $\qty{0.5}{\milli\meter}$ diameter and two chemically-etched needles separated by $\qty{2.8}{\milli\meter}$, which act as the end caps. 
Typically, $V_{DC} = 7-\qty{9}{\volt}$ are applied to the needles, providing axial confinement of the ions. 
The voltage applied to the rods is $|V_{RF}| \approx \qty{230}{\volt}$ delivered at $\nu_{RF} = \qty{20.772}{\mega\hertz}$ by a helical resonator \cite{Low2019}. 

There are three relevant beam paths as depicted in Figure \ref{fig:Fig1}b.
The first beam path is comprised of linearly polarized 493, 554, and $\qty{650}{\nano\meter}$ light focused to beam radii of $\approx$ 31, 35, $\qty{41}{\micro\meter}$, respectively.
This path provides Doppler cooling via $\qty{493}{\nano\meter}$ using the $\ce{6s\:{}^2S_{1/2}} \leftrightarrow \ce{6p\:{}^2P_{1/2}}$ transition, with $\qty{650}{\nano\meter}$ light used to repump $\ce{5d\:{}^2D_{3/2}} \rightarrow \ce{6p\:{}^2P_{1/2}}$.
The $\qty{554}{\nano\meter}$ laser is used to drive the first step in the photoionization process.
The usual laser powers are $110 \pm 3$, $15 \pm 1$, and $210\pm \qty{6}{\micro\watt}$ for the 493, 554, and $\qty{650}{\nano\meter}$ beams, respectively.

The second beam path consists of     linearly polarized 390 or $\qty{405}{\nano\meter}$ light, as well as $\qty{614}{\nano\meter}$ light. 
This path intersects with the first beam path at the center of the trap, $\approx \qty{0.707}{\milli\meter}$ from the trap rods. 
During any given experiment either $\qty{390}{\nano\meter}$ or $\qty{405}{\nano\meter}$ was used as the second photoionization step, but never both. 
In $\mathrm{Ba}^+$, the $\ce{5d\:{}^2D_{5/2}}$ state is populated within a few seconds of exposure to the $\qty{390}{\nano\meter}$ laser, because the $\qty{390}{\nano\meter}$ beam is detuned by $<\qty{1}{\nano\meter}$ from the $\ce{6p\:{}^2P_{1/2}} \rightarrow \ce{6d\:{}^2D_{3/2}}$ transition. 
This can decay to $\ce{5d\:{}^2D_{5/2}}$, so the $\qty{614}{\nano\meter}$ laser is used to repump $\ce{5d\:{}^2D_{5/2}} \rightarrow \ce{6p\:{}^2P_{3/2}}$. 
The $\qty{614}{\nano\meter}$ beam carries $7 \pm \qty{1}{\micro\watt}$ of power. 
It should be noted that the $\qty{614}{\nano\meter}$ is not necessary if the $\qty{390}{\nano\meter}$ beam is shuttered shortly after trapping.
In this set of experiments, the $\qty{390}{\nano\meter}$ was not shuttered so the $\qty{614}{\nano\meter}$ was required for efficient cooling and ion detection. 
The beam radii were $\approx \qty{34}{\micro\meter}$ and $\approx \qty{35}{\micro\meter}$ for the $\qty{390}{\nano\meter}$ and $\qty{405}{\nano\meter}$ beams, respectively. 
The powers of the $\qty{390}{\nano\meter}$ and $\qty{405}{\nano\meter}$ beams vary depending on the experiment, but we only observe trapping above $\qty{150}{\micro\watt}$ in this setup using either laser. 

As shown in Fig. \ref{fig:Fig1}b, the third and final beam consists of only the $\qty{532}{\nano\meter}$ ablation laser. 
This is a nanosecond pulsed, flash-lamp pumped, Nd:YAG laser, capable of delivering up to $\qty{10}{\milli\joule}$ of pulse energy, although we operate in practice at much lower pulse energies, around $\qty{140}{\micro\joule}$. 
Upon hitting the ablation target, this laser pulse causes sublimation of the $\mathrm{BaCl}_2$, generating a vapor of atomic barium.
The ablation target is oriented such that the surface norm is directed towards the trap center, $\qty{14.6}{\milli\meter}$ away \cite{PhysRevA.105.033102}, with the atomic plume reaching the trap within $10-\qty{15}{\micro\second}$. 
The target contains $40-\qty{50}{\milli\gram}$ worth of $\mathrm{BaCl}_2$. 
    
The $\qty{405}{\nano\meter}$ laser is a fiber-pigtailed Cobolt 06-MLD from HÜBNER Photonics GmbH. 
The $\qty{390}{\nano\meter}$ laser is a tunable Littrow laser from MOGLabs. 
In practice, neither of these lasers were frequency locked, although the $\qty{390}{\nano\meter}$ has the ability to be locked using a wavemeter.
The frequencies of the first step for each isotope have already been measured in previous works \cite{PhysRevA.105.033102}. 
For trapping $^{138}\mathrm{Ba}^+$, the $\qty{554}{\nano\meter}$ laser is locked to $\qty{553.70185}{\nano\meter}$. 
The optimal wavelength of the second step is calculated to be $\qty{389.74779}{\nano\meter}$ based on the first step and the previously measured photoionization cross-sections \cite{PCamus_1983}, where the autoionizing resonance occurs $\qty{5.42032}{eV}$ above the $\ce{6s^2\:{}^1S_0}$ state. 
The resonance is $> \qty{50}{\giga\hertz}$ wide, which exceeds the mode-hop free tuning range of the $\qty{390}{\nano\meter}$ Littrow laser, so no attempt was made to rigorously quantify the loading rate as a function of frequency.
In these experiments, the 390 nm laser was simply tuned to $389.748\pm\qty{0.01}{\nano\meter}$, and a significant increase in the loading rate of $^{138}\mathrm{Ba}^+$ was observed, which is detailed in the following sections.

For ablation loading of the ion trap, the following experimental procedure is employed:
\begin{enumerate}
    \item The RF and DC trap voltages are turned on and the laser frequencies are set for $^{138}\mathrm{Ba}^+$.
    \item The ablation laser sends a single pulse to the target. Approximately, $10-\qty{15}{\micro\second}$ later, the majority of the atomic flux reaches the trapping region where the atoms either pass through or are photoionized by the two-step process.
    \item Doppler cooling sweeps are implemented in order to cool and crystallize hot ions that may have been trapped. 
    This is done by red-detuning the $\qty{493}{\nano\meter}$ cooling beam $100-\qty{500}{\mega\hertz}$ and sweeping within five seconds back to the nominal cooling frequency. 
    \item Steps (i)-(iii) are repeated until an ion is trapped. 
\end{enumerate}

    Fluorescent light from neutral atoms and ions is collected by a 0.26 NA home-built objective and sent to either a photo-multiplier tube (H10682-210) or a charge-coupled device (BFLY-PGE-05S2M-CS). 
Typically, experiments were run with light being sent to the photo-multiplier tube to quickly determine whether an ion had been trapped. 
To count how many ions were trapped, the collected light is sent to the charge-coupled device to image the ion chain, as shown in Fig. \ref{fig:Fig1}c.
\section{Results}

\subsection{Saturation Intensity \& Fluence Experiments}

\begin{figure}
    \centering
    \includegraphics[width=\linewidth]{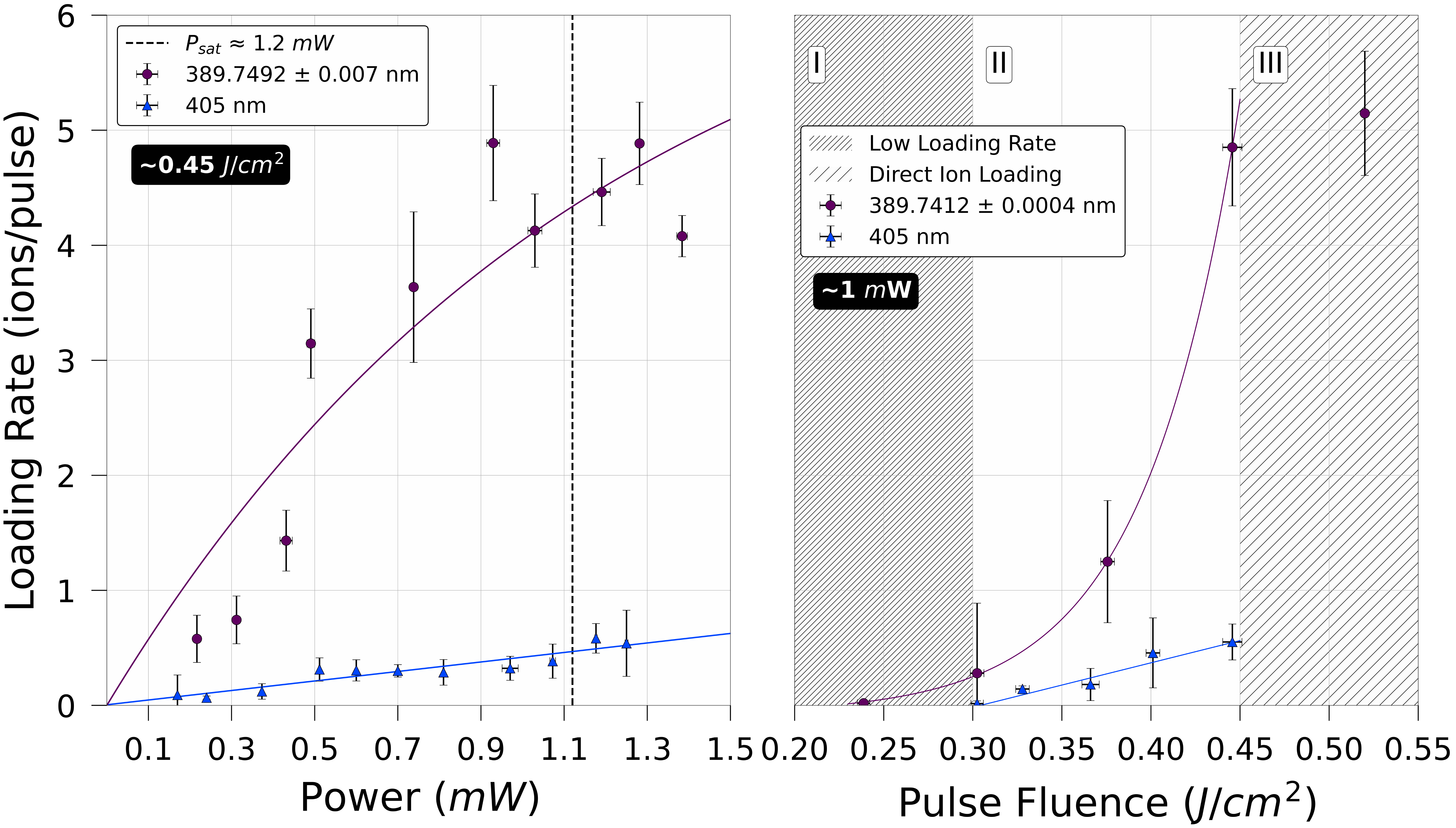}
    \caption{
    a) Loading rates of $^{138}\mathrm{Ba}^+$ as a function of power carried by the second step laser in the two-step photoionization process. 
    The calculated $P_{sat}$ of the autoionizing transition is denoted by the dashed line. 
    The fit to the autoionizing transition assumes the loading rate follows an exponential of the type $f(x) = a(1 - e^{-b\cdot x})$, where $a = 7.10\pm0.06$ ions/pulse and $b = 0.84\pm\qty{0.04}{\per\milli\watt}$ are fitted constants and $x$ is power in milliwatts \cite{PhysRevLett.67.2131}. 
    The fit predicts $f(1/b)\approx 4.5$, which agrees well with the calculated saturation parameter $f(P_{sat})\approx 4.4$. 
    The fit to the non-resonant $\qty{405}{\nano\meter}$ loading rate is linear. 
    The error bars represent the standard error of the mean. 
    The ablation fluence was $\approx\qty{0.45}{\joule\per\centi\meter\squared}$.
    b) Loading rates of $^{138}\mathrm{Ba}^+$ as a function of ablation fluence. 
    We operate primarily in Region II. 
    The regions are described further in the text.
    The power carried by the second step photoionization beam was $\approx\qty{1}{\milli\watt}$. 
    The fitted lines act simply as visual guides.}
    \label{fig:Fig2}
\end{figure}

The following section compares the loading rates utilizing the two photoionization schemes as the second step power is varied. 
We find that the loading rates saturate in the autoionizing scheme at much lower powers compared to the non-resonant scheme. 
Similarly, we vary the ablation fluence and find that the autoionizing scheme can load ions with significantly lower neutral flux. 
Typically, we work with laser powers on the second-step photoionization beam of up to a milliwatt. 
In this regime, the photoionization cross-section in non-resonant schemes is so small that the loading rate scales linearly.
In comparison, autoionizing schemes have quite a large photoionization cross-section with similar amounts of power and the ionization efficiency can reach near 100$\%$ \cite{Payne1985}. 

The loading rates of $^{138}\mathrm{Ba}^+$ (ions/pulse) as a function of laser powers for both photoionization schemes are depicted in Fig. \ref{fig:Fig2}a. 
Below $\qty{200}{\micro\watt}$ the autoionizing and non-resonant photoionization schemes converge to a loading rate of zero.
But at higher powers, the autoionizing transition saturates and the loading rate utilizing the autoionizing transition significantly outperforms that of the non-resonant scheme. 
Saturation of the autoionizing transition occurs at relatively low laser power $\approx \qty{1.2}{\milli\watt}$ with a loading rate of roughly 5 ions/pulse.
Based on a linear fit, the non-resonant scheme is predicted to require at least $\approx \qty{12}{\milli\watt}$ of power to achieve a similar loading rate with a similar beam size as the $\qty{390}{\nano\meter}$.

The saturation power $P_{sat}$ for the autoionizing transition is calculated using the fitted Fano linewidth parameter $\Gamma \approx 60.4\pm\qty{1}{\giga\hertz}$ \cite{BWillke_1993} and saturation intensity $I_{sat} = \dfrac{\hbar\omega^3\Gamma}{4\pi c^2} \approx \qty{63.7}{\watt\per\centi\meter\squared}$, where $c$ is the speed of light and $\omega = 2\pi\cdot\qty{769.211}{\tera\hertz}$ is the angular frequency of the light.
With our beam radius of $w_0 = \qty{34}{\micro\meter}$ for the $\qty{390}{\nano\meter}$ beam, the saturation power $P_{sat}$ is calculated to be:
\begin{equation}
    P_{sat} = \frac{I_{sat}w_0^2\pi}{2}\approx \qty{1.2}{\milli\watt}.
\end{equation}
We were able to achieve saturation on this transition, as shown in Figure \ref{fig:Fig2}a, and our results agree well with the calculated saturation.
The ablation laser pulse fluence was kept constant throughout the saturation experiments at $0.45 \pm \qty{0.01}{\joule\per\centi\meter\squared}$.

Fig. \ref{fig:Fig2}b. shows the loading rate of $^{138}\mathrm{Ba}^+$ as a function of ablation laser fluence. 
The ablation laser fluence experiments were conducted with $1.02\pm 0.02$ and $1.20\pm \qty{0.04}{\milli\watt}$ of average power on the $\qty{405}{\nano\meter}$ and $\qty{390}{\nano\meter}$ beams, respectively.
As the laser fluence increases so does the amount of neutral atoms and ions produced. 
If the fluence is too low, loading can be difficult, but if it is too high, ions can be trapped directly \cite{PhysRevA.105.033102}. 
There are three main regions considered when conducting the ablation loading experiments. 
Region I ($< \qty{0.3}{\joule\per\centi\meter\squared}$) is a region of low laser fluence in which there is minimal loading of $^{138}\mathrm{Ba}^+$ ions. 
We were not able to load in this region using the non-resonant scheme after 100 attempts; however, we were able to successfully load after 53 attempts using the autoionizing scheme.
Region II ($\qty{0.3}-\qty{0.45}{\joule\per\centi\meter\squared}$) is the ideal operating fluence for trapping, since it produces enough neutral atom flux to trap in an isotope-selective manner, but not enough ablation-produced ions to load directly. 
Region III ($> \qty{0.45}{\joule\per\centi\meter\squared}$) is a region of high fluence which produces a significant amount of ions from the ablation process itself, drastically reducing isotope-selectivity since ions can be loaded directly into the trap. 
These regions are not hard limits, but act mostly as guidelines for these experiments conducted with a $\mathrm{BaCl}_2$ ablation target.

\subsection{Optimized Loading Rates}

\begin{figure}
    \centering
    \includegraphics[width=0.9\linewidth]{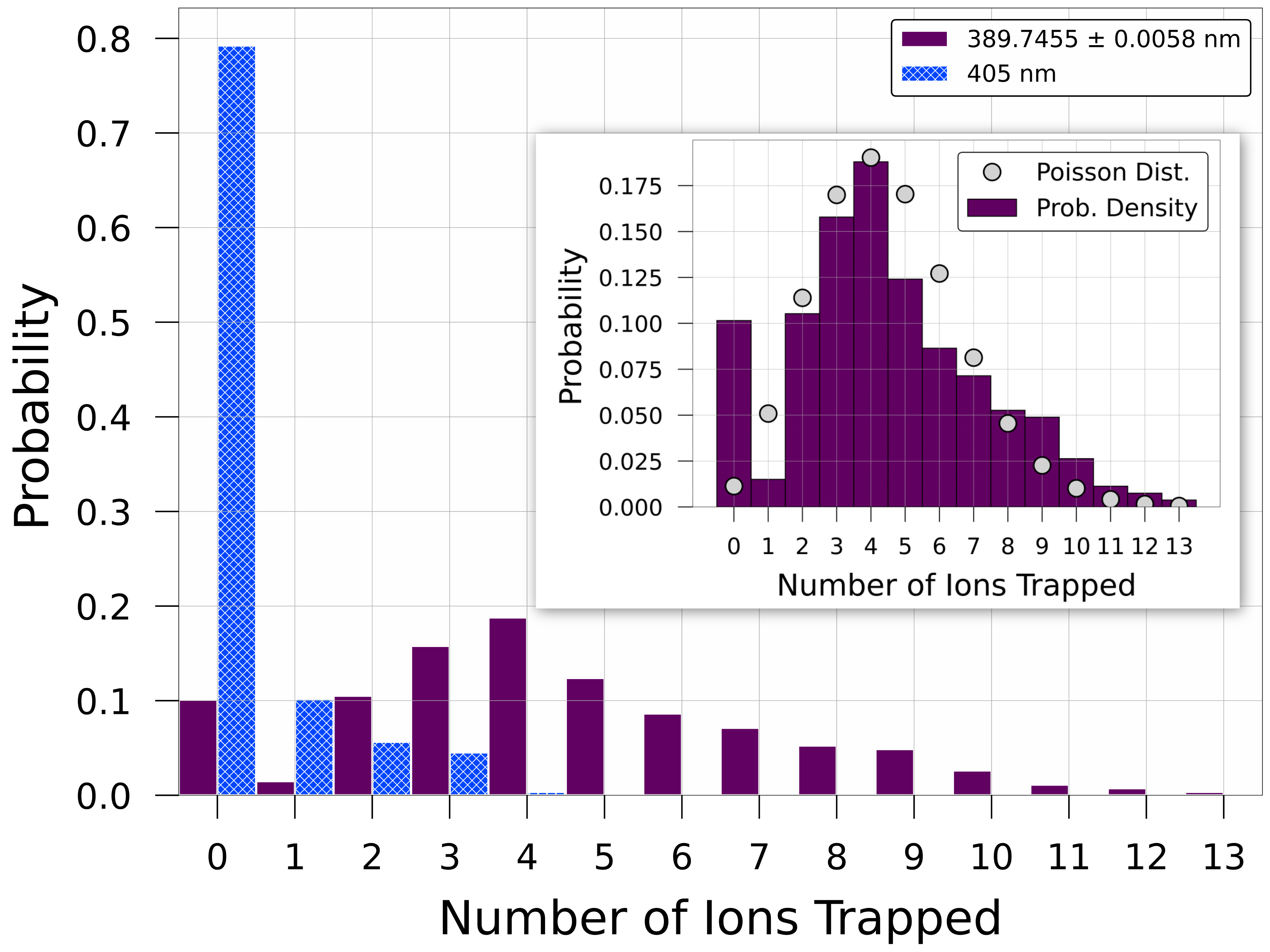}
    \caption{
    Histogram of the loading rates of $^{138}\mathrm{Ba}^+$, where each event is a single ablation pulse and trapping attempt. 
    The average loading rate using the autoionizing scheme was $R_{390} = 4.48\pm0.17$ ions/pulse and the average loading rate using the non-resonant scheme was $R_{405} = 0.43\pm0.11$ ions/pulse. 
    There were 265 and 266 trapping attempts for the non-resonant and autoionizing scheme, respectively. 
    The inset shows the probability of loading using the autoionizing scheme fitted to a Poisson distribution.
    }
    \label{fig:Fig3}
\end{figure}

In this section, we compare the loading rates of $^{138}\mathrm{Ba}^+$ with ideal laser parameters. 
For these experiments, the ablation fluence was set to $\qty{0.45}{\joule\per\centi\meter\squared}$ and the average powers of the second step 405 and $\qty{390}{\nano\meter}$ photoionizing beams were $P_{390} = 1.08\pm 0.01$ and $P_{405} = 1.17\pm \qty{0.01}{\milli\watt}$, respectively. 

The likelihood of trapping long chains of $^{138}\mathrm{Ba}^+$ was greatly increased using the autoionizing scheme, which is evident by the increase in the average ions per pulse. 
Roughly one out of five attempts succeeded in trapping using the non-resonant scheme, while nine out of ten attempts succeeded using the autoionizing scheme. 
The average loading rates with the standard error of the mean are $R_{405} = 0.43\pm0.11$ ions/pulse and $R_{390} = 4.48\pm0.17$ ions/pulse using the non-resonant and autoionizing schemes, respectively.
These rates are compared in Fig. \ref{fig:Fig3}.
The median ions trapped per pulse was 0 ions for the non-resonant scheme and 4 ions for the autoionizing scheme.

The loading rates quoted above cannot be directly compared to infer the enhancement from the autoionizing resonance, because the powers of the second-step lasers and the quantity of neutral atoms produced during ablation were not identical across both sets of experiments. 
During the loading experiments, we monitor the rate of fluorescence at $\qty{554}{\nano\meter}$ after each ablation laser pulse, as a proxy for the flux of neutral barium atoms.
We observed higher neutral atom flux for the experiments conducted with the autoionizing scheme than with the non-resonant scheme (up to 39.5$\%$ higher fluorescence counts at $\qty{554}{\nano\meter}$), which we attribute to natural variations in the number of ablated atoms. 
We make the simplifying assumptions that the probability of trapping is proportional to both the neutral atom flux and the average second-step laser power, when the ionization rate is far from saturation \cite{PhysRevA.105.033102}.
Thus, we estimate that the loading rate using $\qty{405}{\nano\meter}$ would have been $R'_{405} = 0.66\pm0.24$ ions/pulse, if the neutral atom flux and second-step laser power had been identical to those in the $\qty{390}{\nano\meter}$ experiments.
We therefore calculate that the increase in loading rate of $^{138}\mathrm{Ba}^+$ using the autoionizing scheme is a factor of $R_{390}/R'_{405}\approx 6.8\pm2.7$.
This compares well with the ratio of photoionization cross-sections, which have previously been measured as  $\sigma_{390}/\sigma_{405} = \qty{520}{\mega\barn}$/$\qty{75}{\mega\barn}$ $\approx 6.9$ \cite{BWillke_1993} and $\qty{550}{\mega\barn}$/$\qty{60}{\mega\barn}$ $\approx 9.2$ \cite{Kalyar}.
\section{Discussion}

The demonstrated autoionizing scheme using 554 and $\qty{390}{\nano\meter}$ retains the isotope-selectivity previously presented in \cite{PhysRevA.105.033102}, since it utilizes the same first step, which is the isotope-selective step in the photoionization process. 
We conducted a brief study to confirm this, where $478$ ions (minimum of two $^{138}\mathrm{Ba}^+$ ions in the chain, a total of 114 unique ion chains) were analyzed for dark or dim ions, which indicate the presence of other ion isotopes besides $^{138}\mathrm{Ba}^+$. 
We found that only 8 ions in all of these chains were other isotopes. 
This puts a maximum bound on the isotope-selectivity for $^{138}\mathrm{Ba}^+$ of 98$\%$, which is consistent with previous studies \cite{PhysRevA.105.033102}.
To achieve high loading rates, the first transition in the ionization sequence is usually power broadened significantly, which limits isotope selectivity \cite{PhysRevA.105.033102}. 
Thus, using this more efficient autoionizing scheme and lowering the power of the first step laser may allow for better isotope selectivity with similar loading rates in the future.

We briefly investigated the loading rates of the $^{137}\mathrm{Ba}^+$ isotope with the autoionizing scheme.
A significant increase in the loading rate, with a factor of at least 5, was immediately apparent for the autoionizing scheme.
However, we did not make a controlled quantitative comparison between the two schemes for this isotope.
This photoionization scheme can also be applied to $^{133}\mathrm{Ba}^+$ since the autoionizing transition is broad enough that it makes any isotope shifts between $^{138}\mathrm{Ba}^+$, $^{137}\mathrm{Ba}^+$, and $^{133}\mathrm{Ba}^+$ negligible. 

The linewidth of the measured autoionizing resonance is $\approx \qty{60}{\giga\hertz}$, but we found a considerable increase in the loading rate even when the frequency of the $\qty{390}{\nano\meter}$ laser was $> \qty{100}{\giga\hertz}$ detuned from the $\qty{389.74779}{\nano\meter}$ transition.
So in practice, tuning anywhere near this resonance can result in a significant increase in the loading rate. 
In fact, the autoionizing transition is so broad that precision control of the laser frequency and linewidth are not required.

There are many autoionizing resonances embedded in the first ionization continuum of barium that can be used for efficient photoionization. 
Assuming $\qty{554}{\nano\meter}$ is the first step in a two-step photoionization scheme, the best second step is to utilize the strongest autoionizing resonance, as this will most efficiently ionize the neutral atoms and allow for smaller amounts of optical power to be used. 
Other possible autoionizing states that are easily accessible from the intermediate $\ce{6s 6p\:{}^1P_1}$ state are outlined in Table \ref{table:OtherResonances}. 

\begin{table}
\begin{center}
    \begin{tabular}{c|ccc}
    \toprule\toprule
    Wavelength (nm) & Cross-section (Mb) & Autoion. State \cite{PCamus_1983} & Energy (eV) \cite{BWillke_1993} \\ \hline
    380.75 & $\approx 300$ \cite{BWillke_1993} & \ce{5d_{3/2} 9d} $(J = 0)$& 5.49549 \\
    384.33 & $\approx 290$ \cite{BWillke_1993} & \ce{5d_{3/2} 9d} $(J = 1)$& 5.46516\\ 
    389.74 & 550 \cite{BWillke_1993}, $520 \pm 78$ \cite{Kalyar}& \ce{5d_{5/2} 8d} $(J = 1)$& 5.42032 \\ 
    402.92 & 370 \cite{BWillke_1993}, $300 \pm 45$ \cite{Kalyar}& \ce{5d_{3/2} 8d} $(J = 1)$& 5.31629 \\ \bottomrule\bottomrule
    \end{tabular}
    \caption{
    Comparison of some previously measured autoionizing resonances when using $\ce{6s 6p\:{}^1P_1}$ as the intermediate state.
    Other intermediate states can be used, but the highest experimentally measured photoionization cross-section of barium utilizes $\ce{6s 6p\:{}^1P_1}$ as the intermediate state. 
    The energy is relative to the $\ce{6s^2\:{}^1S_0}$ ground state.
    }
    \label{table:OtherResonances}
\end{center}
\end{table}

Other first steps can be used besides $\qty{554}{\nano\meter}$ to drive to an intermediate state before exciting an autoionizing state embedded in the continuum. 
This includes, but is not limited to, the \ce{5d 6p\:{}^3D_1} state driven with $\qty{413}{\nano\meter}$ light \cite{PhysRevA.47.1981} and the \ce{6s 6p\:{}^3P_1} state driven with $\qty{791}{\nano\meter}$ light \cite{Kalyar}.
The highest resonance in the photoionization cross-section utilizing the \ce{6s 6p\:{}^3P_1} intermediate state has been measured to be $102 \pm \qty{15}{\mega\barn}$ using $\qty{340}{\nano\meter}$ as the second step \cite{Kalyar}. This is a significantly strong resonance, but still considerably lower than the $\qty{550}{\mega\barn}$ cross-section using \ce{6s 6p\:{}^1P_1} as the intermediate state with $\qty{390}{\nano\meter}$. 
There are considerable resonances in the photoionization spectrum of barium using \ce{5d 6p\:{}^3D_1} as an intermediate state, for which relative cross-sections have been measured \cite{PhysRevA.47.1981}.


\section{Conclusion}
We have demonstrated an efficient isotope-selective photoionization scheme for trapping barium ions utilizing an autoionizing resonance, motivated by the elusive $^{133}\mathrm{Ba}^+$ and $^{137}\mathrm{Ba}^+$ isotopes.
The increase in loading rates (ions/pulse) compared with previous ablation studies are consistent with previously measured photoionization cross-sections of barium and equate to nearly an order of magnitude increase in the loading rate of $^{138}\mathrm{Ba}^+$ ions. 
This photoionization scheme will aid in the consistent trapping of long chains of barium ions and will help enable NISQ era devices to be built based on rare isotopes like $^{133}\mathrm{Ba}^+$. 
Furthermore, the demonstrated advantage of utilizing autoionizing resonances allows researchers to efficiently ionize neutral atoms without having to focus significant amounts of power near the trap, reducing the risk of trap charging. 

\section*{Acknowledgments}
This research was supported in part by the Natural Sciences and Engineering Research Council of Canada (NSERC) and the Canada First Research Excellence Fund (CFREF). We would like to thank Dr. Jens Lassen, a research scientist at TRIUMF, for his initial suggestion and for pointing out the existence of these autoionizing resonances as well as ongoing discussion related to the experiment. We thank Dr. Akbar Jahangiri Jozani, Nicholas Zutt, and Xinghe Tan for reviewing the manuscript.
\section*{Conflict of interest}
The authors declare no conflicts of interest.
\newline
\bibliography{references}

\providecommand{\noopsort}[1]{}\providecommand{\singleletter}[1]{#1}%
\begin{thebibliography}{10}

\bibitem{omg}
D.~T.~C. Allcock, W.~C. Campbell, J.~Chiaverini, I.~L. Chuang, E.~R. Hudson,
  I.~D. Moore, A.~Ransford, C.~Roman, J.~M. Sage, and D.~J. Wineland.
\newblock omg blueprint for trapped ion quantum computing with metastable
  states.
\newblock {\em Applied Physics Letters}, 119(21), 11 2021.
\newblock 214002.

\bibitem{low2023control}
Pei~Jiang Low, Brendan White, and Crystal Senko.
\newblock Control and readout of a 13-level trapped ion qudit, 2023.

\bibitem{PhysRevA.41.2621}
A.~A. Madej and J.~D. Sankey.
\newblock Quantum jumps and the single trapped barium ion: Determination of
  collisional quenching rates for the 5${d}^{2}$${D}_{5/2}$ level.
\newblock {\em Phys. Rev. A}, 41:2621--2630, 3 1990.

\bibitem{binaimotlagh2023guided}
Ali Binai-Motlagh, Matthew Day, Nikolay Videnov, Noah Greenberg, Crystal Senko,
  and Rajibul Islam.
\newblock A guided light system for agile individual addressing of ba$^+$
  qubits with $10^{-4}$ level intensity crosstalk, 2023.

\bibitem{Hucul2017}
D.~Hucul, J.~E. Christensen, Eric~R. Hudson, and Wesley~C. Campbell.
\newblock Spectroscopy of a synthetic trapped ion qubit.
\newblock {\em Phys. Rev. Lett.}, 119:100501, 9 2017.

\bibitem{Christensen2020_thesis}
J.~E. Christensen.
\newblock {\em High-fidelity operation of a radioactive trapped ion qubit,
  $^{133}\mathrm{Ba}$}.
\newblock PhD thesis, University of California, Los Angeles, 2020.

\bibitem{PhysRevLett.129.130501}
Fangzhao~Alex An, Anthony Ransford, Andrew Schaffer, Lucas~R. Sletten, John
  Gaebler, James Hostetter, and Grahame Vittorini.
\newblock High fidelity state preparation and measurement of ion hyperfine
  qubits with $i>\frac{1}{2}$.
\newblock {\em Phys. Rev. Lett.}, 129:130501, 9 2022.

\bibitem{shi2023ablation}
X.~Shi, S.~L. Todaro, G.~L. Mintzer, C.~D. Bruzewicz, J.~Chiaverini, and I.~L.
  Chuang.
\newblock Ablation loading of barium ions into a surface electrode trap, 2023.

\bibitem{Graham2014}
R.~D. Graham, S.-P. Chen, T.~Sakrejda, J.~Wright, Z.~Zhou, and B.~B. Blinov.
\newblock A system for trapping barium ions in a microfabricated surface trap.
\newblock {\em AIP Advances}, 4(5):057124, 2014.

\bibitem{ASHFOLD19991424}
Michael~N.R. Ashfold and Colin~M. Western.
\newblock Multiphoton spectroscopy, applications.
\newblock In John~C. Lindon, editor, {\em Encyclopedia of Spectroscopy and
  Spectrometry}, pages 1424--1433. Elsevier, Oxford, 1999.

\bibitem{Niggli1987}
S.~Niggli and M.~C. Huber.
\newblock {Transition probabilities in neutral barium}.
\newblock {\em Phys. Rev. A}, 35(7):2908--2918, 1987.

\bibitem{AKULSHIN199254}
A.M. Akulshin, A.A. Celikov, and V.L. Velichansky.
\newblock Nonlinear doppler-free spectroscopy of the 61s0-63p1 intercombination
  transition in barium.
\newblock {\em Optics Communications}, 93(1):54--58, 1992.

\bibitem{PhysRevA.61.062509}
V.~A. Dzuba, V.~V. Flambaum, and J.~S.~M. Ginges.
\newblock Calculation of parity and time invariance violation in the radium
  atom.
\newblock {\em Phys. Rev. A}, 61:062509, May 2000.

\bibitem{PhysRev.38.873}
A.~G. Shenstone.
\newblock Ultra-ionization potentials in mercury vapor.
\newblock {\em Phys. Rev.}, 38:873--875, 9 1931.

\bibitem{vant2006photoionization}
K.~Vant, J.~Chiaverini, W.~Lybarger, and D.~J. Berkeland.
\newblock Photoionization of strontium for trapped-ion quantum information
  processing, 2006.

\bibitem{PhysRevA.74.063421}
L.~Deslauriers, M.~Acton, B.~B. Blinov, K.-A. Brickman, P.~C. Haljan, W.~K.
  Hensinger, D.~Hucul, S.~Katnik, R.~N. Kohn, P.~J. Lee, M.~J. Madsen,
  P.~Maunz, S.~Olmschenk, D.~L. Moehring, D.~Stick, J.~Sterk, M.~Yeo, K.~C.
  Younge, and C.~Monroe.
\newblock Efficient photoionization loading of trapped ions with ultrafast
  pulses.
\newblock {\em Phys. Rev. A}, 74:063421, 12 2006.

\bibitem{UFano_1983}
U~Fano.
\newblock Correlations of two excited electrons.
\newblock {\em Reports on Progress in Physics}, 46(2):97, 2 1983.

\bibitem{PhysRev.124.1866}
U.~Fano.
\newblock Effects of configuration interaction on intensities and phase shifts.
\newblock {\em Phys. Rev.}, 124:1866--1878, 12 1961.

\bibitem{BWillke_1993}
B~Willke and M~Kock.
\newblock Measurement of photoionization cross sections from the laser-excited
  ba i (6s6p) 1p10 state.
\newblock {\em Journal of Physics B: Atomic, Molecular and Optical Physics},
  26(6):1129, 3 1993.

\bibitem{M_Aymar_1990}
M~Aymar.
\newblock Eigenchannel r-matrix calculation of the j=1 odd-parity spectrum of
  barium.
\newblock {\em Journal of Physics B: Atomic, Molecular and Optical Physics},
  23(16):2697, 8 1990.

\bibitem{MAymar_1979}
M~Aymar and O~Robaux.
\newblock Multichannel quantum-defect analysis of the bound even-parity j=2
  spectrum of neutral barium.
\newblock {\em Journal of Physics B: Atomic and Molecular Physics}, 12(4):531,
  2 1979.

\bibitem{RevModPhys.68.1015}
Mireille Aymar, Chris~H. Greene, and Eliane Luc-Koenig.
\newblock Multichannel rydberg spectroscopy of complex atoms.
\newblock {\em Rev. Mod. Phys.}, 68:1015--1123, 10 1996.

\bibitem{PhysRevA.42.5773}
Chris~H. Greene and Constantine~E. Theodosiou.
\newblock Photoionization of the ba 6s6p $^{1}$${\mathit{p}}_{1}$ state.
\newblock {\em Phys. Rev. A}, 42:5773--5775, 11 1990.

\bibitem{PhysRevA.44.1773}
Chris~H. Greene and Mireille Aymar.
\newblock Spin-orbit effects in the heavy alkaline-earth atoms.
\newblock {\em Phys. Rev. A}, 44:1773--1790, 8 1991.

\bibitem{PhysRevLett.67.2131}
L.-W. He, C.~E. Burkhardt, M.~Ciocca, J.~J. Leventhal, and S.~T. Manson.
\newblock Absolute cross sections for the photoionization of the 6s6p $^{1}$p
  excited state of barium.
\newblock {\em Phys. Rev. Lett.}, 67:2131--2134, 10 1991.

\bibitem{Kalyar}
M.~M. Ali~Kalyar.
\newblock {\em Two-step Laser Excitation Studies of Bound and Autoionizing
  States in Barium}.
\newblock PhD thesis, Quaid-i-Azam University, Islamabad, 2008.

\bibitem{osti_1326630}
Peter Maunz, Jonathan Mizrahi, and Josh Goldberg.
\newblock Ioncontrol v. 1.0, version 00, 6 2016.

\bibitem{PhysRevA.105.033102}
Brendan~M. White, Pei~Jiang Low, Yvette de~Sereville, Matthew~L. Day, Noah
  Greenberg, Richard Rademacher, and Crystal Senko.
\newblock Isotope-selective laser ablation ion-trap loading of
  $^{137}\mathrm{Ba}^{+}$ using a ${\mathrm{bacl}}_{2}$ target.
\newblock {\em Phys. Rev. A}, 105:033102, 3 2022.

\bibitem{Low2019}
P.~J. Low.
\newblock Tolerable experimental imperfections for a quadrupole blade ion trap
  and practical qudit gates with trapped ions.
\newblock Master's thesis, University of Waterloo, 2019.

\bibitem{PCamus_1983}
P~Camus, M~Dieulin, A~El Himdy, and M~Aymar.
\newblock Two-step optogalvanic spectroscopy of neutral barium: Observation and
  interpretation of the even levels above the 6s ionization limit between 5.2
  and 7 ev.
\newblock {\em Physica Scripta}, 27(2):125, 2 1983.

\bibitem{Payne1985}
M.~G. Payne and G.~S. Hurst.
\newblock {\em Theory of Resonance Ionization Spectroscopy}, pages 183--188.
\newblock Springer US, Boston, MA, 1985.

\bibitem{PhysRevA.47.1981}
Darrell~J. Armstrong, Robert~P. Wood, and Chris~H. Greene.
\newblock Photoionization of the 5d6p $^{3}$${\mathit{d}}_{1}$ state of barium.
\newblock {\em Phys. Rev. A}, 47:1981--1988, Mar 1993.

\end{thebibliography}
\bibliographystyle{unsrt}

\end{document}


\title{Supplementary}
\section{Calculation of Rabi Frequency}
Here we discuss how the two-photon Rabi rate in figure 3 (b) of the main text was calculated. Taking into account coupling to both the $P_{1/2}$ and $P_{3/2}$ excited states, it can be shown that the Rabi rate is given by \cite{Ozeri2007, Wineland2003}:
\begin{equation}
    \Omega = \frac{g_bg_r}{3}(b_-r_--b_+r_+)\frac{\omega_f}{\Delta(\Delta-\omega_f)}
    \label{eq:RamanRabi}
\end{equation}
where $\omega_f$ is the fine structure splitting between the $P_{1/2}$ and $P_{3/2}$ levels, $\Delta$ is the detuning from the $P_{1/2}$ level and $b_-, r_-, b_+, r_+$ are the left and right circular polarization components of the red (r) and blue (b) detuned beams involved in the raman process. By picking the polarization of the two beams to be linear but orthogonal ($b_-=r_-=b_+=-r_+=1/\sqrt{2}$), we can maximize the Rabi rate.  Finally $g_{b/r} = E_{b/r}|\bra{P_{3/2}, F=I+3/2, m_F=F}\boldsymbol{d}\cdot \hat{\sigma}_+\ket{S_{1/2}, F=I+1/2, m_F=F}|/2\hbar$ where $E_{b/r}$ is the electric field amplitude at the position of the ions. This can be related to the optical power using the relationship between electric field and intensity, $E=\sqrt{\frac{2I}{c\epsilon_0}}$, and the relationship between power and the peak intensity of a gaussian beam, $I = \frac{2P}{\pi \omega_x\omega_y}$. Here $\omega_x$ and $\omega_y$ are the beam radii along the two principal axes of an elliptical beam. Using the Wigner-Eckart theorem one can show that the above matrix element is equivalent to the reduced matrix element $\bra{J=3/2}|e\boldsymbol{r}|\ket{J^\prime=1/2}$ \cite{Steck}. The lifetime of the $P_{3/2}$ state can then be used to compute this quantity using \cite{Steck}:
\begin{equation}
    \frac{1}{\tau} = \frac{8\pi^2}{3\epsilon_0\hbar\lambda^3}\frac{2J+1}{2J^\prime+1} |\bra{J}|e\boldsymbol{r}|\ket{J^\prime}|^2
\end{equation}
where $\lambda = 455$ nm is the wavelength of the transition and $\tau$ is the lifetime of this excited state.

\begin{table}[H]
\centering
\begin{tabular}{l|l}
Parameter                     & \begin{tabular}[c]{@{}l@{}}Value \\\end{tabular}  \\ 
\hline
IA and global Intensity                   & 4.2$\times 10^4$ W/cm$^2$    \\                                                                             
$\Delta$ & -280 THz
\\
$\omega_f$ & $2 \pi\times 51$ THz\\
$\tau$ & 7.9 ns
\end{tabular}
\end{table}

\section{Microlens array data}

This section includes data on the characterization of the full grid of microlenses. Each row in this grid was designed to have the same effective focal length (EFL). The EFLs of the rows start from 0.525 mm and go up to 1 mm in steps of 0.025 mm for a total of 20 rows. The decision to manufacture lens arrays with different EFLs was made to account for manufacturing tolerances.  To determine whether a row of microlenses has the required EFL, an optical profilometer was used to measure the surface profile of one lens from each row. Cross-sections of these surface profile measurements are shown in figure \ref{fig:OpticalProfile} (a). From this data, the radius of curvature of each lens, as well as the EFL can be extracted. This data is shown in table \ref{table:EFL_Data}. Figure \ref{fig:OpticalProfile} (b) shows the deviation of the radius of curvature of the manufactured lenses from the design. On average the manufactured lenses have an EFL that is 2\% shorter than the design.

\begin{figure}[H]
    \centering
    \includegraphics[scale=0.65]{MLACharacterization.pdf}
    \caption{(a) Cross-sections of the surface profiles of the manufactured lenses obtained from an optical profilometer. (b) Normalized difference between the radii of curvature of the designed and manufactured lenses. The manufactured lenses have a radius of curvature that is on average 2\% shorter than the design, as indicated by the green line.}
    \label{fig:OpticalProfile}
\end{figure}

\begin{table}[H]
\centering
\begin{tabular}{l|l|l|l}
MLA~ & Design Curvature (mm)~ & Manufactured lens Curvature (mm)~ & EFL (mm)~  \\ 
\hline
1~   & -0.15422~              & -0.1524~                          & 0.5156~    \\
2~   & -0.15877~              & -0.1549~                          & 0.5289~    \\
3~   & -0.1631~               & -0.1609~                          & 0.5621~    \\
4~   & -0.1673~               & -0.1652~                          & 0.5867~    \\
5~   & -0.1714~               & -0.1655~                          & 0.5885~    \\
6~   & -0.1753~               & -0.1710~                          & 0.6209~    \\
7~   & -0.1790~               & -0.1731~                          & 0.6336~    \\
8~   & -0.1826~               & -0.1870~                          & 0.7249~    \\
9~   & -0.1861~               & -0.1820~                          & 0.6910~    \\
10~  & -0.1894~               & -0.1822~                          & 0.6925~    \\
11~  & -0.1927~               & -0.1896~                          & 0.7432~    \\
12~  & -0.1958~               & -0.1919~                          & 0.7592~    \\
13~  & -0.1989~               & -0.1935~                          & 0.7708~    \\
14~  & -0.2018~               & -0.1973~                          & 0.7995~    \\
15~  & -0.2047~               & -0.2033~                          & 0.8456~    \\
16~  & -0.2076~               & -0.2070~                          & 0.8760~    \\
17~  & -0.2104~               & -0.2059~                          & 0.8669~    \\
18~  & -0.2132~               & -0.2087~                          & 0.8898~    \\
19~  & -0.2160~               & -0.2109~                          & 0.9085~    \\
20~  & -0.2189~               & -0.2100~                          & 0.9015~   
\end{tabular}
\caption{Radii of curvature of the design and manufactured lenses, along with the effective focal lengths of the fabricated lenses.}
\label{table:EFL_Data}
\end{table}